\documentclass[a4paper]{article}

\usepackage{icrc2013}
\usepackage[english]{babel}

\title{The NectarCAM camera project}

\shorttitle{The NectarCAM camera project}

\authors{
J-F.Glicenstein$^{1}$,
M.Barcelo$^{11}$,
J-A. Barrio$^{12}$,
O.Blanch$^{11}$,
J.Boix$^{11}$,
J.Bolmont$^{4}$,
C.Boutonnet$^{2}$
S.Cazaux$^{1}$,
E.Chabanne$^{7}$,
C.Champion$^{2}$,
F.Chateau$^{1}$,
S.Colonges$^{2}$,
P.Corona$^{4}$,
S.Couturier$^{5}$,
B.Courty$^{2}$,
E.Delagnes$^{1}$,
C.Delgado$^{10}$,
J-P.Ernenwein$^{6}$,
S.Fegan$^{5}$,
O.Ferreira$^{5}$,
M.Fesquet$^{1}$,
G.Fontaine$^{5}$,
N.Fouque$^{7}$,
F.Henault$^8$,
D.Gasc\'on$^{13}$,
D.Herranz$^{12}$,
R.Hermel$^{7}$,
D.Hoffmann$^{6}$,
J.Houles$^{6}$,
S.Karkar$^{4}$,
B.Khelifi$^{5}$,
J.Kn\"odlseder$^{3}$,
G.Martinez$^{10}$,
K.Lacombe$^{3}$,
G.Lamanna$^{7}$,
T.LeFlour$^{7}$,
R.Lopez-Coto$^{11}$,
F.Louis$^{1}$,
A.Mathieu$^{5}$,
E.Moulin$^{1}$,
P.Nayman$^{4}$,
F.Nunio$^{1}$,
J-F. Olive$^{3}$,
J-L. Panazol$^{7}$,
P-O. Petrucci$^{8}$,
M.Punch$^{2}$,
J.Prast$^{7}$,
P.Ramon$^{3}$,
M.Riallot$^{1}$,
M.Rib\'o$^{13}$,
S.Rosier-Lees$^{7}$,
A.Sanuy$^{13}$,
J.Siero$^{13}$,
J-P.Tavernet$^{4}$,
L.A.Tejedor$^{12}$,
F.Toussenel$^{455}$,
G.Vasileiadis$^{9}$,
V.Voisin$^{4}$,
V.Waegebert$^{3}$,
C.Zurbach$^{9}$,
for the CTA consortium.
}

\afiliations{
$^1$ DSM/IRFU, CEA-Saclay, F91191 Gif-sur-Yvette, France \\
$^2$ APC, AstroParticule et Cosmologie, Universite Paris Diderot, CNRS/IN2P3, CEA/Irfu, Observatoire de Paris, Sorbonne Paris Cite, 10, rue Alice Domon et Leonie Duquet, 75205 Paris Cedex 13, France \\
$^3$ Institut de Recherche en Astrophysique et Planetologie, 9 av. colonel Roche, BP 44 346, 31028 Toulouse Cedex 4, France \\
$^4$ LPNHE, Universite Pierre et Marie Curie Paris 6, Universite Denis Diderot Paris 7, CNRS/IN2P3, 4 Place Jussieu, F-75252, Paris Cedex 5, France \\ 
$^5$ Laboratoire Leprince-Ringuet, Ecole Polytechnique, CNRS/IN2P3, F-91128 Palaiseau, France \\
$^6$ Centre de Physique des Particules de Marseille, CNRS/IN2P3, 168 Avenue de Luminy, 13009 Marseille, France  \\
$^7$ LAPP, Universite de Savoie, CNRS/IN2P3, 9 Chemin de Bellevue - BP 110, 74941 Annecy-le-Vieux Cedex, France \\
$^8$ UJF-Grenoble 1 CNRS/INSU, Institut de Planetologie et d'Astrophysique de Grenoble (IPAG)  UMR 5274, Grenoble, F-38041, France \\
$^9$ Laboratoire Univers et Particules de Montpellier, Universite Montpellier 2, CNRS/IN2P3, CC 72, Place Eugene Bataillon, F-34095 Montpellier Cedex 5, France\\
$^{10}$ CIEMAT, Avda. Complutense 22, 28040 Madrid, Spain \\ 
$^{11}$ Institut de Fisica d'Altes Energies (IFAE) Edifici Cn, Campus UAB, 08193 Bellaterra, Spain \\
$^{12}$ Grupo de Altas Energias, Universidad Complutense de Madrid, Av Complutenses, 28040 Madrid, Spain \\
$^{13}$ Departament d'Astronomia i Meteorologia (ICC-UB), Universitat de Barcelona, Marti i Franques, 1, 08028, Barcelona, Spain  
}

\email{glicens@cea.fr}

\abstract{In the framework of the next generation of Cherenkov telescopes, the Cherenkov Telescope Array (CTA), NectarCAM is a camera designed for the medium size telescopes covering the central energy range of 100 GeV to 30 TeV. NectarCAM will be finely pixelated ($\sim$ 1800 pixels for a 8$^o$ field of view, FoV) in order to image atmospheric Cherenkov showers by measuring the charge deposited within a few nanoseconds time-window. It will have additional features like the capacity to record the full waveform with GHz sampling for every pixel and to measure event times with nanosecond accuracy. An array of a few tens of medium size telescopes, equipped with NectarCAMs, will achieve up to a factor of ten improvement in sensitivity over existing instruments in the energy range of 100 GeV to 10 TeV. The camera is made of roughly 250 independent read-out modules, each composed of seven photo-multipliers, with their associated high voltage base and control, a read-out board and a multi-service backplane board. The read-out boards use NECTAr (New Electronics for the Cherenkov Telescope Array) ASICs which have the dual functionality of analogue memories and Analogue to Digital Converter (ADC). The camera trigger to be used will be flexible so as to minimize the read-out dead-time of the NECTAr chips. We present the camera concept and the design and tests of the various subcomponents. The design includes the mechanical parts, the cooling of the electronics, the readout, the data acquisition, the trigger, the monitoring and services.}

\keywords{Methods, techniques and instrumentation}

\begin{document}
\maketitle

\section{Introduction}
NectarCAM is a new camera design for the medium size telescopes (MST) of the planned CTA\footnote{http://www.cta-observatory.org} array of Imaging Atmospheric Cerenkov telescopes (IACTs). The CTA will have several sizes of single or dual-mirror telescopes. The single-mirror MST has a 12-meter diameter dish.  
Its camera will face new challenges:
\begin{enumerate}
\item The fields of view of camera are expected to be larger (e.g. at least 7$^o$ compared to 5$^o$ for the cameras of H.E.S.S.-1, an previous generation array of IACTs). As a consequence more distant showers will be seen in the cameras. The arrival of the photons from a distant shower can last more than 100 nanoseconds. However, the signal to noise ratio in a pixel is optimized by recording and later integrating the signal in a much smaller window ($\sim$ 10 ns). If the camera trigger is not flexible enough (e.g  is similar to the H.E.S.S. camera trigger), the event energies would be systematically underestimated. Special trigger strategies have been designed to trigger different pixels at different moments (see Figure 16 of \cite{bib:nectar}). These strategies still need to be tested. The trigger rate is also expected to increase by a factor of two compared to the previous generation of Cherenkov telescopes, as result of the larger field of view, potetially leading to a larger dead time of the instrument. 
\item Large timing gradients are expected in high-energy gamma-ray events. Measuring these gradients improves the reconstruction of the events. Analyzing the whole waveform also helps improve the hadron rejection (see e.g., Aliu et al. \cite{bib:Aliu}). It has been estimated that a convenient reconstruction of the time gradient could be achieved with a time resolution of 2 ns per pixel. To achieve a proper time resolution, it may be necessary to analyze the whole waveform sampled at 1 GHz. In the H.E.S.S. camera, an average charge, time and time spread per pixel was calculated on-board for every triggered event. Sending the whole waveform of the event to the ground implies a data rate per event roughly 5 times larger than that of H.E.S.S. Taking into account the increase in trigger rate, the data rate will increase by an order of magnitude compared to H.E.S.S.  

\item To achieve the aforementionned 2 ns time resolution per pixel, the systematics on the time measurement such as the analogue bandwidth of the read-out system, or the delay between pixels due to the transit time spread of phototubes have to be  carefully evaluated and corrected whenever possible.
\end{enumerate}
As a result of the larger field of view and better timing capabilities of the future Cherenkov cameras, it will be possible to improve the reconstruction of the events and to have a better hadron background rejection. 
However, these new capabilities present new challenges for Cherenkov telescopes.

\begin{figure}[!t]
\centering
\includegraphics[width=8cm]{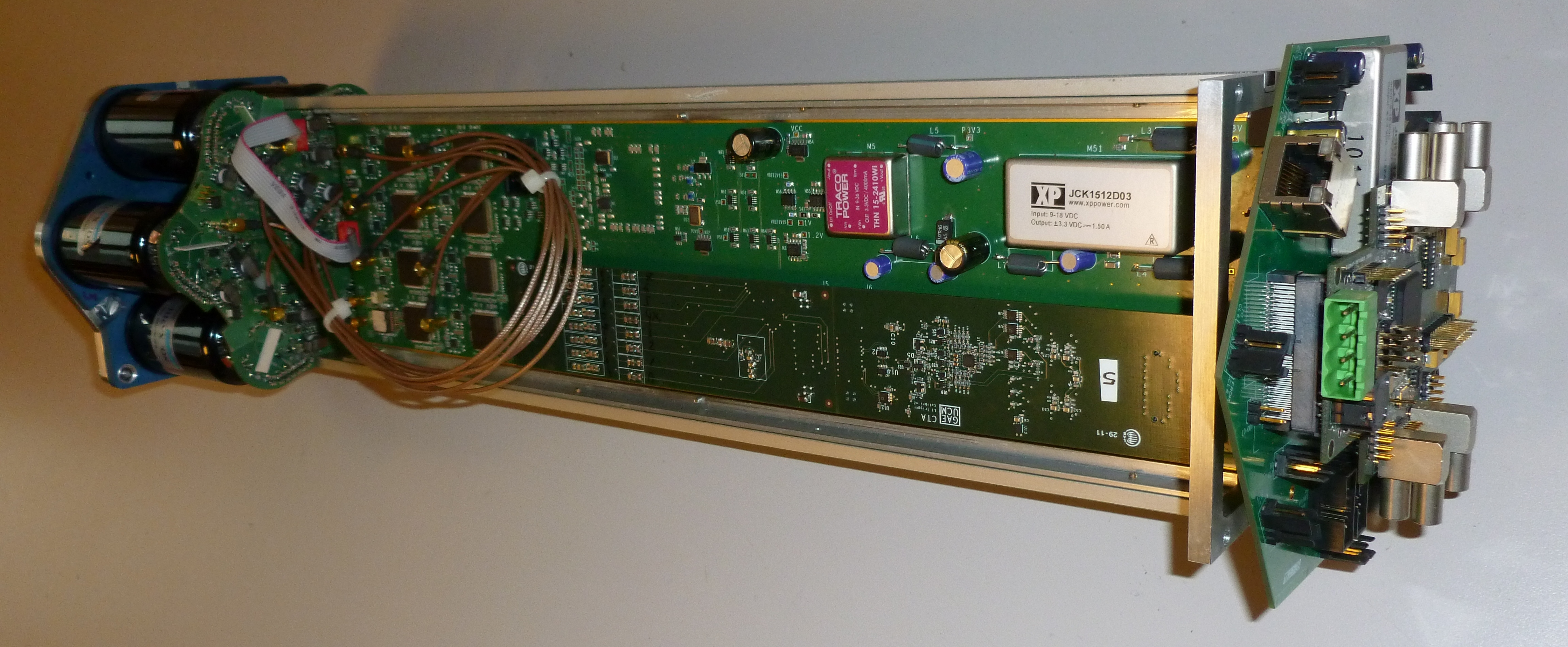}
\caption{Early version (V0) of the NECTAr readout module used in the NectarCAM camera.}
\label{fig:NECTAr-module}
\end{figure}

Two starting points for the NectarCAM are the architecture of the H.E.S.S.-2 camera and a read-out module: the NECTAr module \cite{bib:nectar}. Figure
\ref{fig:NECTAr-module} shows an early version of the NECTAr module.
This module converts the light of a set of photomultiplier tubes (PMTs) into a digital signal when some triggering conditions are fulfilled. 
The modular architecture of the H.E.S.S.-2 camera has the advantage of 
avoiding costly cables and allowing for easy maintenance.
The modular mechanical structure is described in Section \ref{sec:mechanics}. For the NectarCAM, modules are groups of seven pixels, with the associated front-end board, which are completely autonomous in what concerns the power and control electronics. The only cables that go to the ground are the input power voltage and optical fibers for network communications and array trigger connections. 
Preliminary versions of the readout module exist. 
 The photo-sensors, the readout and trigger system, the acquisition system, 
the monitoring and the power supplies are described in 
Sections \ref{sec:photosensors}, 
\ref{sec:readout}, \ref{sec:daq} and \ref{sec:monitoring} respectively. 
The tests of a  camera demonstrator, composed of 7 modules, and the future plans are presented in Section \ref{sec:demonstrator}.

\section{Global structure and mechanics}\label{sec:mechanics} 
\begin{figure}[!t]
\centering
\includegraphics[width=8cm]{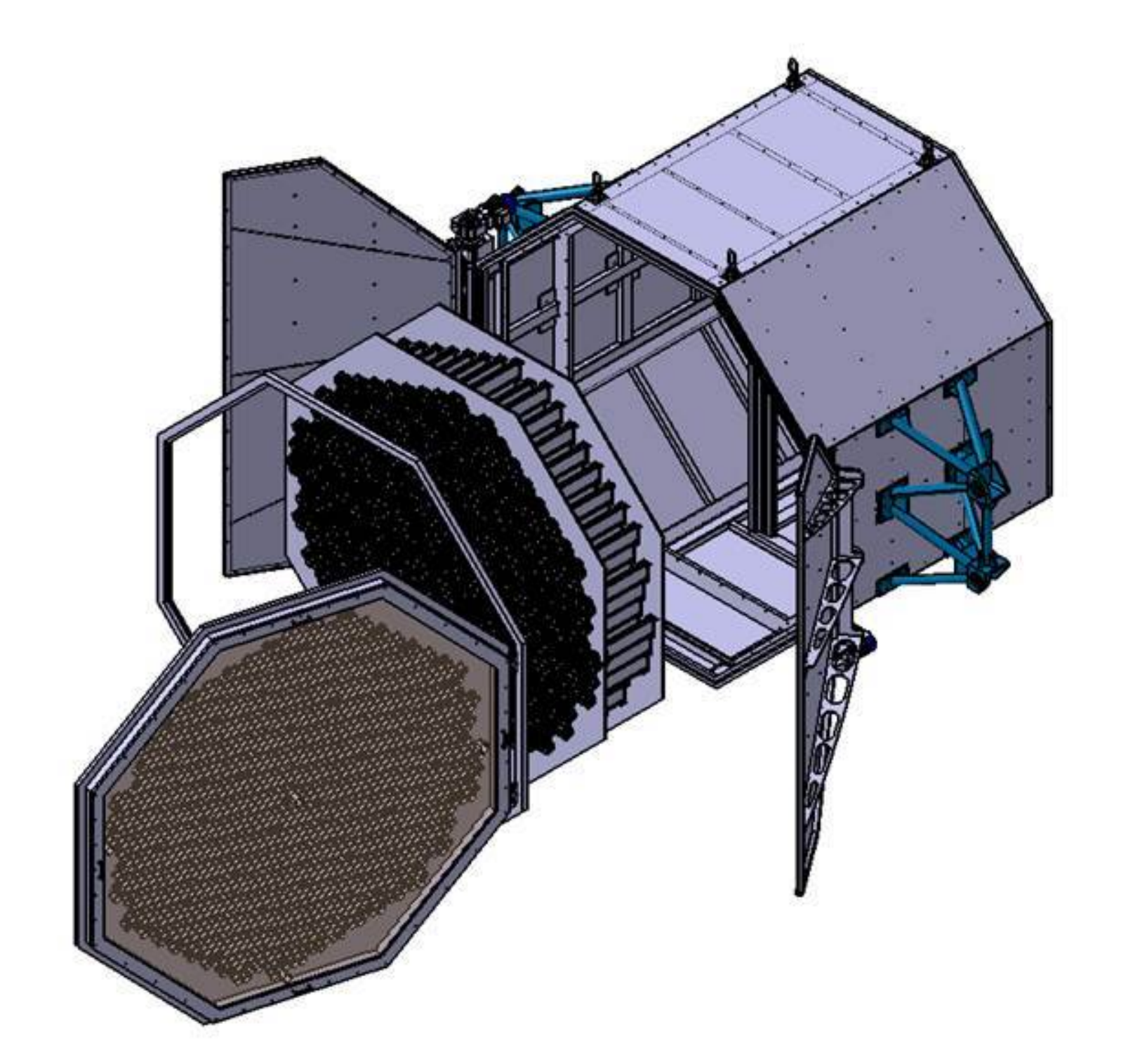}
\caption{Exploded view of the NectarCAM camera mechanics}
\label{fig:camera}
\end{figure}

All the scientific equipment and internal mechanical pieces are contained within a global structure, called the skeleton, that also provides the rigidity for the structure (Figure \ref{fig:camera}). The camera sealing is realized by aluminium honeycomb plates, creating the so-called skin, enclosing the camera equipment. The front part of the camera is sealed against rain by motorized lids controlled either remotely or locally. These lids have a secondary function to hold equipment for the electronics calibration (e.g. a Mylar plate for the single photo-electron calibration) and for the pointing calibration (positioning LEDs, reflecting screen). 
An additional plexiglass plate is placed between the lid and the entrance
of the focal plane to avoid dust contamination.  
The camera will have $\sim 1800$ PMTs grouped in $\sim 250$ modules.
These modules are inserted into a global structure called the sandwich. The rest of the internal equipment (services, communication and electrical interface with the exterior, cables) are held by mechanical fixtures such as small racks, wiring ducts and electrical cabinets. The weight of the camera is less than 2 tons. The total power consumption is of the order of 7.4 kW.   

\begin{figure}[!t]
\centering
\includegraphics[width=6cm]{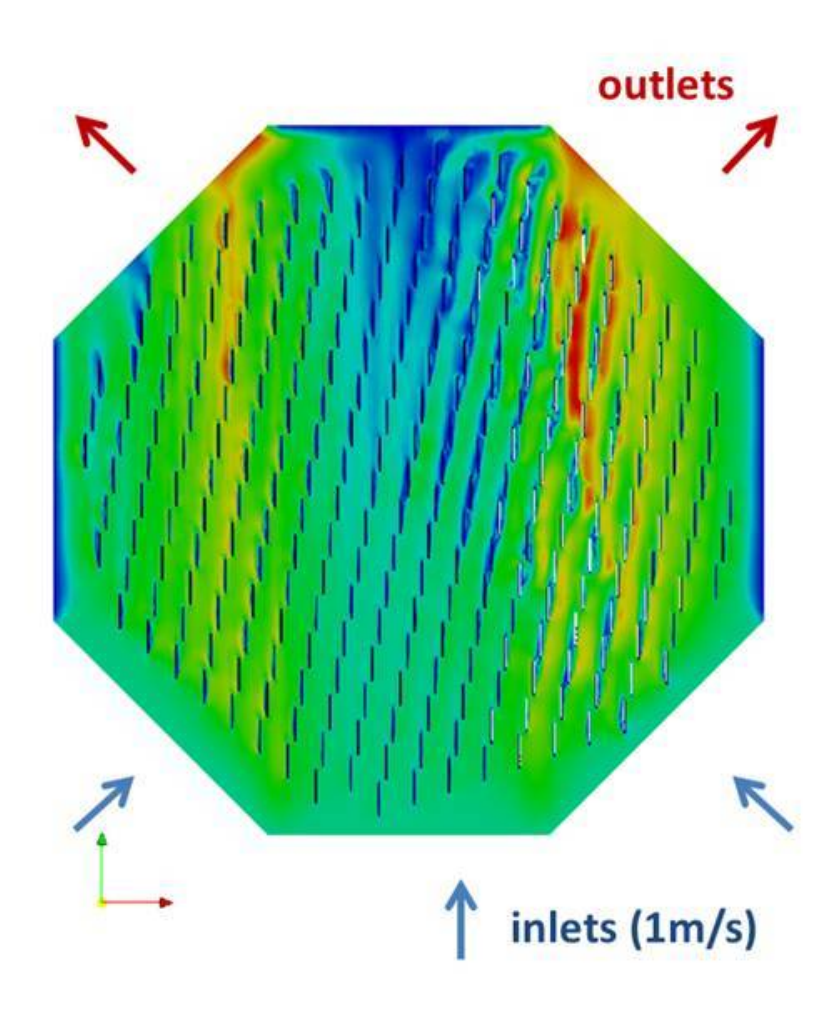}
\caption{Simulation of a cooling airflow in the NectarCAM sandwich. The sandwich hosts $\sim 250$ NECTAr modules.}
\label{fig:cooling}
\end{figure}
Since the camera is sealed, proper care has to be taken to avoid overheating of the camera.
Air-based cooling systems have been studied (Figure \ref{fig:cooling}), and merging these with a water-cooling system is under study.

\section{Photodetector and detector unit}\label{sec:photosensors}
The focal plane of the NectarCAM camera is equipped with {\it detector units}. These detector units are composed of a photo-detector, the associated 
high voltage power supply, and an amplifier. 
CTA has decided to use PMTs  for the photodetectors of their single mirror telescopes. A candidate PMT is the R11920-100 developed by Hamamatsu\footnote{www.hamamatsu.com}. The R11920-100 PMT has a single photon electron signal FWHM of 2.5 to 3 ns. The light from the dead spaces between PMTs will be collected by custom designed Winston cones or lenses. The high voltage of the PMTs will be obtained from an external 24-V supply with an ASIC, similar to a chip designed for the KM3Net experiment \cite{bib:Gajanana}. The output from the PMTs is amplified by a wideband (450 MHz) 16-bit amplifier called PACTA \cite{bib:PACTA}. The large bandwidth of the PACTA allows a minimal distortion of the PMT signal, and thus a better measurement of the arrival time of detected photons. 
\begin{figure}[!t]
\centering
\includegraphics[width=8cm]{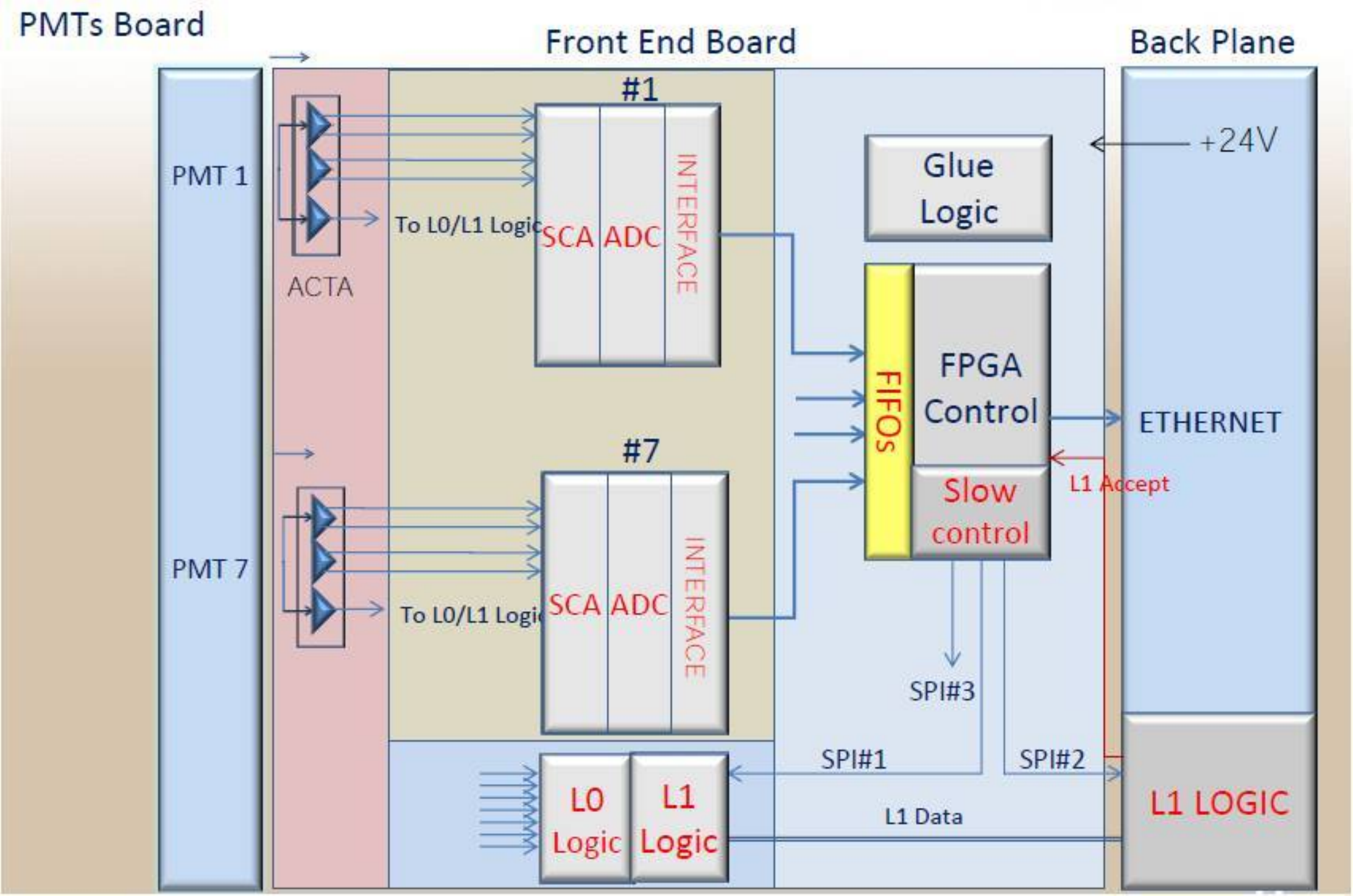}
\caption{Principle of NECTAr readout modules.}
\label{fig:readout}
\end{figure}

\section{Readout}\label{sec:readout}
\begin{figure}[!t]
\centering
\includegraphics[width=8cm]{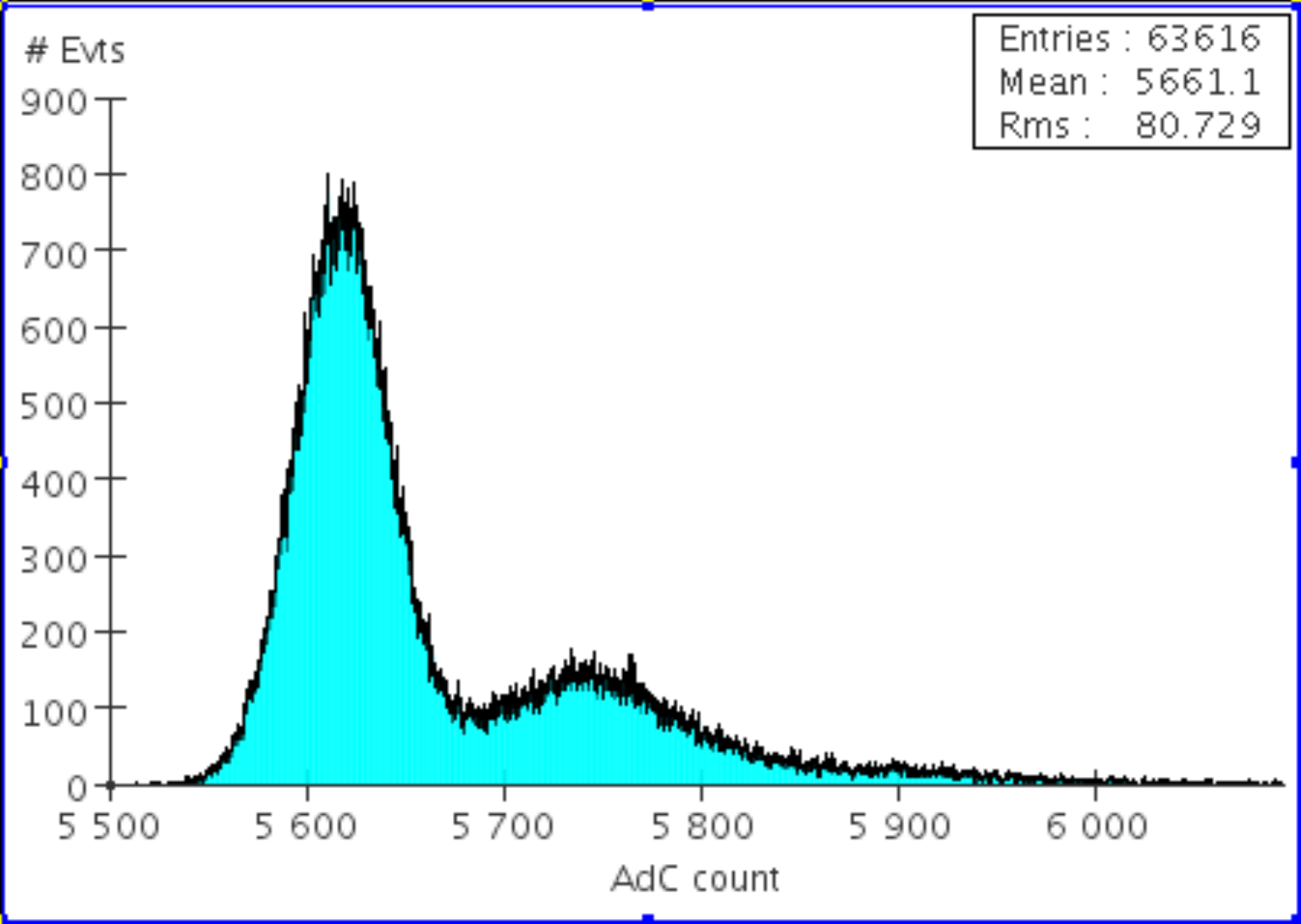}
\caption{Single photoelectron spectrum obtained with the NectarCAM V1 module. 
The PMT signal is amplified by a PACTA and an ACTA amplifier.}
\label{fig:single-pe}
\end{figure}

As illustrated in Figure \ref{fig:readout}, NECTAr modules are composed
of 3 blocks. The block on the left hand side is composed of 7 PMTs, each associated with its detector unit, as described in Section \ref{sec:photosensors}. The central block is the front-end board. Signals from the detector units are amplified again in an ACTA\cite{bib:ACTA} ASIC. At this level, the signal is divided into a low-gain, high-gain (relative gain 16) and a trigger channel. The trigger strategy is described in Section \ref{sec:daq}. The two first levels of the trigger (L0 and L1) are temporarily implemented as mezzanines on the read-out board. The outputs of the low-gain and high-gain channels of the ACTA are sent to a NECTAr chip\cite{bib:nectarchip}. The NECTAr chip has the dual functionality of switched capacitor array and analog-to-digital convertor. The switched capacitor array, which acts as a circular buffer, has a depth of 1024 samples and can be operated between 500 MHz and 3.2 GHz. The NECTAr chip has a bandwidth of more than 400 MHz and a dynamic range of 11.3 bits. The power consumption is 210 mW. The dead time is 2 $\mu$s for the readout of 16 cells. The NECTAr chip is thus a low power, cheap alternative to the use of a Flash Analogue to Digital Convert (FADC). Once a L1 trigger signal is emitted, the data from a 16-20 ns time interval around the date of trigger are read out. They are retrieved from the NECTAr chip and sent to a FPGA located on the front end board. This FPGA has several functions: interface to the ethernet, control of the NECTAr chip, of the L0, L1 trigger configurations, and the HV power supply. High level quantites such as the integrated charge over the readout window and the arrival time of the signal can be calculated inside the FPGA. It is then possible to send these quantities over the ethernet connection instead of the full event sample. The last block, on the right hand side of Figure  \ref{fig:readout} is a backplane board. It is used for the clock and synchronisation signal distribution, the low voltage (24 V) power distribution, and the L0/L1 trigger distribution. In some trigger schemes, one of the backplane boards could be used to make the L1 (camera) trigger decision. The readout board has been tested with photon and electrical signals. Its charge response is linear. It allows the measurement of signals between 1 and 3000 photo-electrons. The single photoelectron peak can be measured for calibration purposes (see Figure \ref{fig:single-pe}).

\section{Data acquisition and trigger}\label{sec:daq}
\begin{figure}[!t]
\centering
\includegraphics[width=8cm]{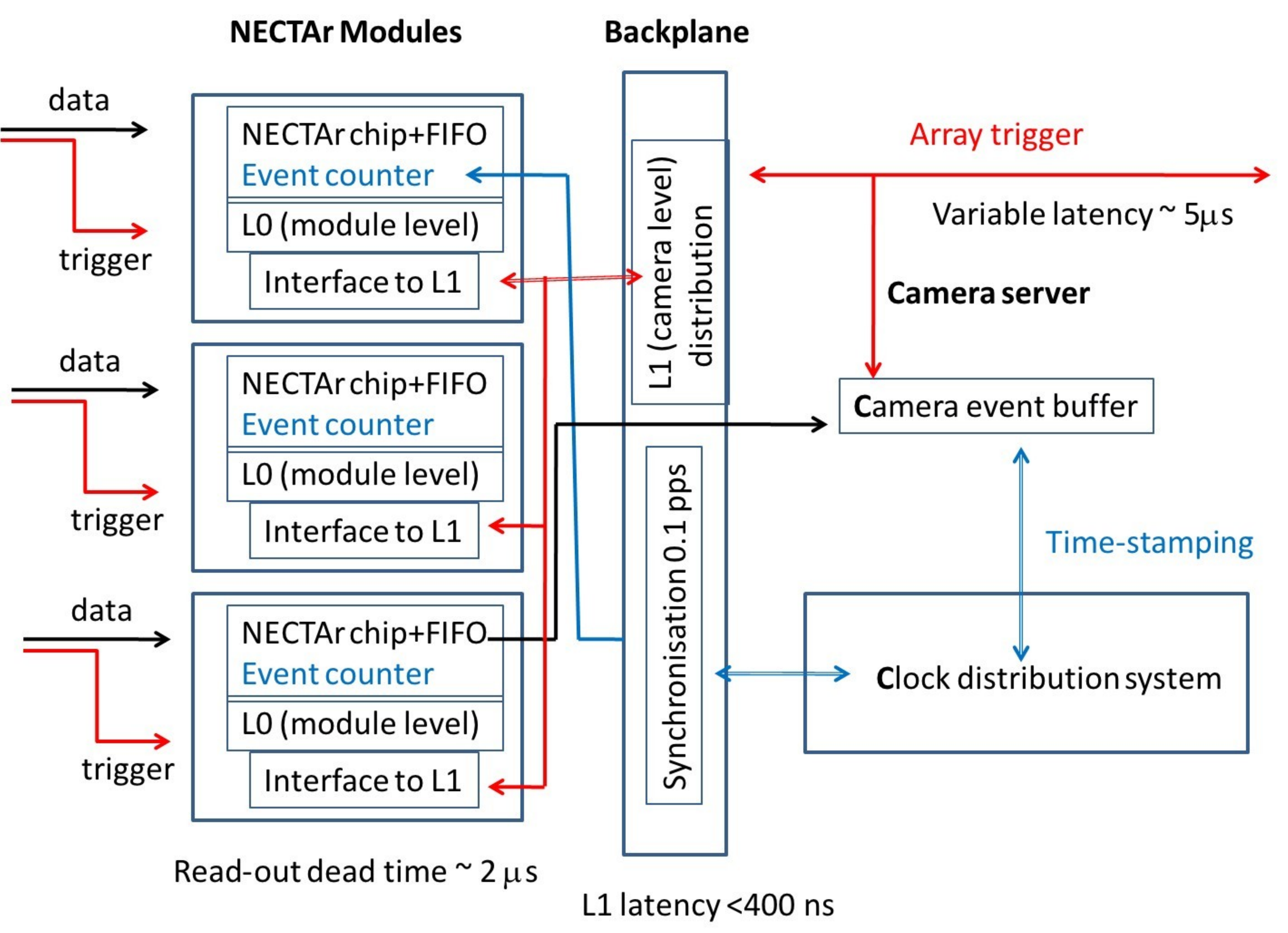}
\caption{Triggering scheme of the NectarCAM camera.}
\label{fig:trigger}
\end{figure}

The NectarCAM camera is triggered in a multilevel scheme, shown in Figure \ref{fig:trigger}. The first level trigger (L0) is a module-level trigger. The information from several modules is combined to realize a camera-level trigger. The L0 and L1 trigger can be implemented with an ``analogue''\cite{bib:analogue} or a ``digital'' solution. The latency of the camera trigger is less than 400 nanoseconds. This is much less than the depth of the switched capacitor array in the NECTAr chip, so that trigged events can be read back from the past. The trigged events are time-stamped and sent to a camera server by ethernet. Events on the camera  server are accepted only if they are coincident with triggers from one or several other telescopes. The typical trigger rate of a single telescope is 5 kHz. The latency of the array trigger is a few $\mu$s. Single telescope events can be time stamped with an accuracy of a few ($\sim 2$) nanoseconds.  
The data acquisition follows the model described in \cite{bib:acquisition}.
Before being sent to the camera server, data from NECTAr modules are concentrated in $\sim 7$ switches. The switches are linked to the camera server through three 10 Gbit connections. The data rate between a NECTAr module and the switches is 2 Mbit/s, if only the total charge and arrival time of pixels are transferred.
It is $\sim$ 40 Mbit/s if all the samples in the region of interest are transferred.

\section{Slow control and services}\label{sec:monitoring}
NectarCAM will have sensors to monitor the temperature, pressure and
humidity inside the camera. Its safety will be ensured with ambient light sensors, smoke detectors and by tracking the position of the lids and back doors. Most subsystems of the camera are controlled remotely. This is the case for the calibration-related hardware (positionning leds), for the cooling system, for the DAQ crate and the clock distribution board. The monitoring/slow control of the NectarCAM will use either industrial solutions such as Programmable Logic 
Controllers, or custom made boards with FPGAs. The latter solution, where the sensors are accessed by industrial buses such as I2C, has been used for the H.E.S.S-2 telescope. 

NectarCAM is powered by industrial low voltage supplies, since the High Voltage for operating the PMTs is created in the detector units 
(Section \ref{sec:photosensors}).

\section{Camera demonstrators}\label{sec:demonstrator}
\begin{figure}[!t]
\centering
\includegraphics[width=8cm]{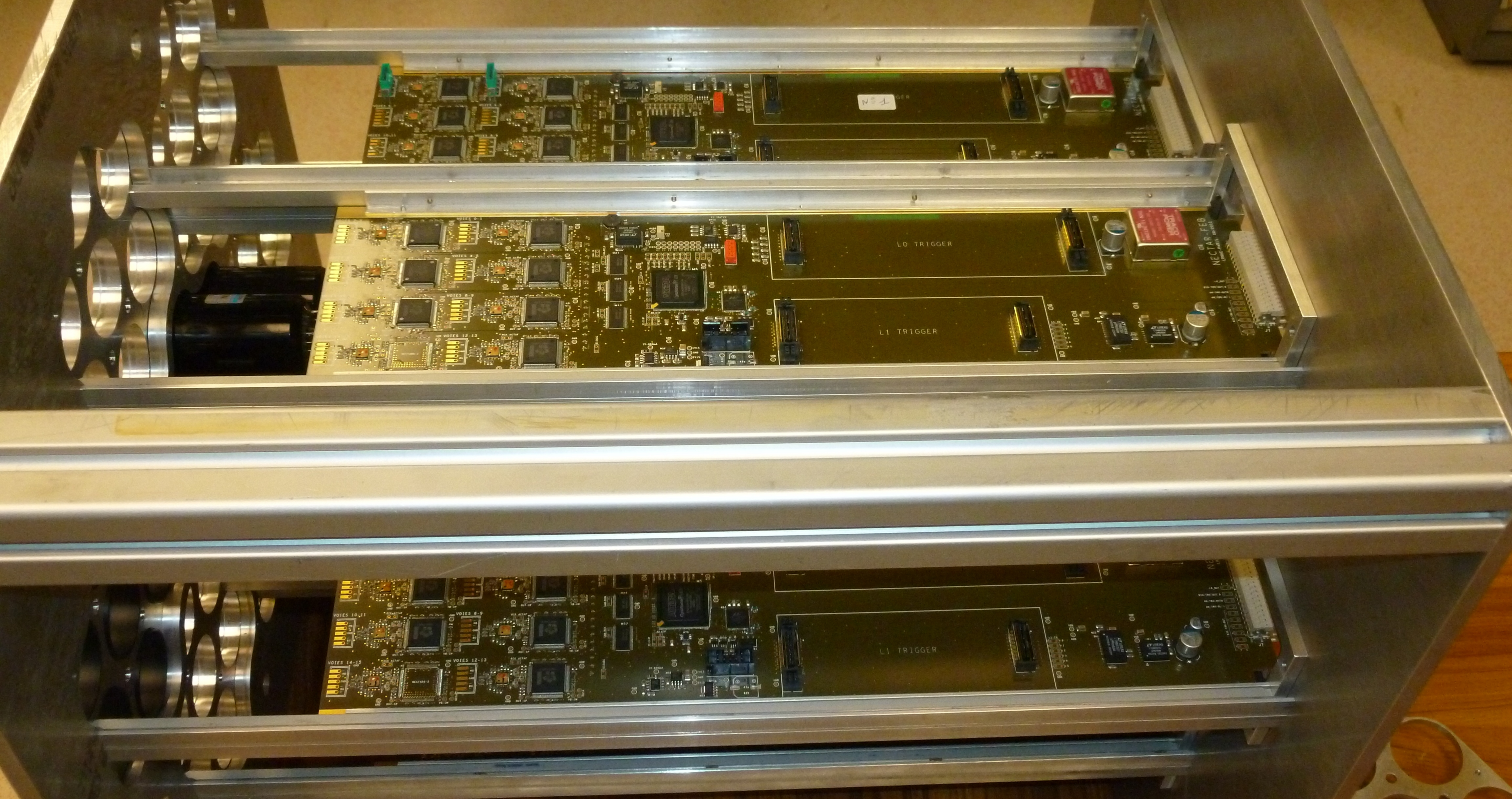}
\caption{A 7-module demonstrator of the NectarCAM camera.}
\label{fig:demonstrator}
\end{figure}

The performance of the NectarCAM, especially the timing and triggering performances, the cooling and data acquisition will be tested with demonstrators. A first demonstrator with 7 NECTAr modules (Figure \ref{fig:demonstrator}) is readily available. It will enable to test the performance of the analogue and digital triggers. The construction of a second demonstrator with 19 modules should start by the end of the year, with the aim of testing the integration of the components and the industrial processes.

\section{Acknowledgements}
\footnotesize{
We gratefully acknowledge support from the agencies and organizations  
  listed in this page http: http://www.cta-observatory.org/?q=node/22.
}







\end{document}